\documentclass[letter, prl, superscriptaddress, twocolumn, floatfix]{revtex4-1}

\usepackage{setspace}
\usepackage{graphicx} 	
\usepackage{dcolumn} 	
\usepackage{bm}   		
\usepackage{amssymb}   	
\usepackage{siunitx}		
\usepackage[version=3]{mhchem}	
\usepackage{miller}
\usepackage{mathtools}
\usepackage{setspace}
\usepackage{color}

\def\k{{\bm k}}
\def\q{{\bm q}}

\begin{document}

\title{Giant negative magnetoresistance driven by spin-orbit coupling at the \ce{LaAlO3}/\ce{SrTiO3} interface}

\author{M.~Diez}
\altaffiliation{These two authors contributed equally to this work.}
\affiliation{Instituut-Lorentz, Universiteit Leiden, P.O. Box 9506, 2300 RA Leiden, The Netherlands}
\author{A.M.R.V.L.~Monteiro}
\altaffiliation{These two authors contributed equally to this work.}
\affiliation{Kavli Institute of Nanoscience, Delft University of Technology, Lorentzweg 1, 2628 CJ Delft, The Netherlands}
\author{G.~Mattoni}
\affiliation{Kavli Institute of Nanoscience, Delft University of Technology, Lorentzweg 1, 2628 CJ Delft, The Netherlands}
\author{E.~Cobanera}
\affiliation{Instituut-Lorentz, Universiteit Leiden, P.O. Box 9506, 2300 RA Leiden, The Netherlands}
\affiliation{Institute for Theoretical Physics, Center for Extreme Matter and Emergent Phenomena, Utrecht University, Leuvenlaan 4, 3584 CE Utrecht, The Netherlands}
\author{T.~Hyart}
\affiliation{Instituut-Lorentz, Universiteit Leiden, P.O. Box 9506, 2300 RA Leiden, The Netherlands}
\affiliation{Department of Physics and Nanoscience Center, University of Jyv\"{a}skyl\"{a}, P.O. Box 35 (YFL), FI--40014 University of Jyv\"{a}skyl\"{a}, Finland}
\author{E.~Mulazimoglu}
\affiliation{Kavli Institute of Nanoscience, Delft University of Technology, Lorentzweg 1, 2628 CJ Delft, The Netherlands}
\author{N.~Bovenzi}
\affiliation{Instituut-Lorentz, Universiteit Leiden, P.O. Box 9506, 2300 RA Leiden, The Netherlands}
\author{C.W.J.~Beenakker}
\affiliation{Instituut-Lorentz, Universiteit Leiden, P.O. Box 9506, 2300 RA Leiden, The Netherlands}
\author{A.D.~Caviglia}
\affiliation{Kavli Institute of Nanoscience, Delft University of Technology, Lorentzweg 1, 2628 CJ Delft, The Netherlands}

\date{January 2015}

\begin{abstract}
The \ce{LaAlO3}/\ce{SrTiO3} interface hosts a two-dimensional electron system that is unusually sensitive to the application of an in-plane magnetic field. Low-temperature experiments have revealed a giant negative magnetoresistance (dropping by 70\%), attributed to a magnetic-field induced transition between interacting phases of conduction electrons with Kondo-screened magnetic impurities. Here we report on experiments over a broad temperature range, showing the persistence of the magnetoresistance up to the 20~K range --- indicative of a single-particle mechanism. Motivated by a striking correspondence between the temperature and carrier density dependence of our magnetoresistance measurements we propose an alternative explanation. Working in the framework of semiclassical Boltzmann transport theory we demonstrate that the combination of spin-orbit coupling and scattering from finite-range impurities can explain the observed magnitude of the negative magnetoresistance, as well as the temperature and electron density dependence.
\end{abstract}

\maketitle

The mobile electrons at the \ce{LaAlO3}/\ce{SrTiO3} (LAO/STO) interface \cite{ohtomoa2004} display an exotic combination of superconductivity \cite{reyren2007,caviglia2008} and magnetic order \cite{brinkman2007,bert2011,li2011,kalisky2012}. The onset of superconductivity at sub-Kelvin temperatures appears in an interval of electron densities where the effect of Rashba spin-orbit coupling on the band structure at the Fermi level is strongest \cite{shalom2010,caviglia2010}, but whether this correlation implies causation remains unclear.

Transport experiments above the superconducting transition temperature have revealed a very large (``giant'') drop in the sheet resistance of the LAO/STO interface upon application of a parallel magnetic field \cite{benshalom2009,Wan11,joshua2012,joshua2013}. An explanation has been proposed \cite{joshua2013,ruhman2014} in terms of the Kondo effect: Variation of the electron density or magnetic field drives a quantum phase transition between a high-resistance correlated electronic phase with screened magnetic impurities and a low-resistance phase of polarized impurity moments. The relevance of spin-orbit coupling for magnetotransport is widely appreciated \cite{benshalom2009, trushin2009, flekser2012, fete2012, ruhman2014, Caprara2012, Bucheli2014}, but it was generally believed to be too weak an effect to provide a single-particle explanation of the giant magnetoresistance.

In this work we provide experimental data (combining magnetic field, gate voltage, and temperature profiles for the resistance of the LAO/STO interface) and theoretical calculations that support an explanation fully within the single-particle context of Boltzmann transport. The key ingredients are the combination of spin-orbit coupling, band anisotropy, and finite-range electrostatic impurity scattering. The thermal insensitivity of the giant magnetoresistance \cite{benshalom2009, Wan11}, in combination with a striking correspondence that we have observed between the gate voltage and temperature dependence of the effect, are features that are difficult to reconcile with the thermally fragile Kondo interpretation --- but fit naturally in the semiclassical Boltzmann description.

\begin{figure}
\includegraphics[width=\linewidth]{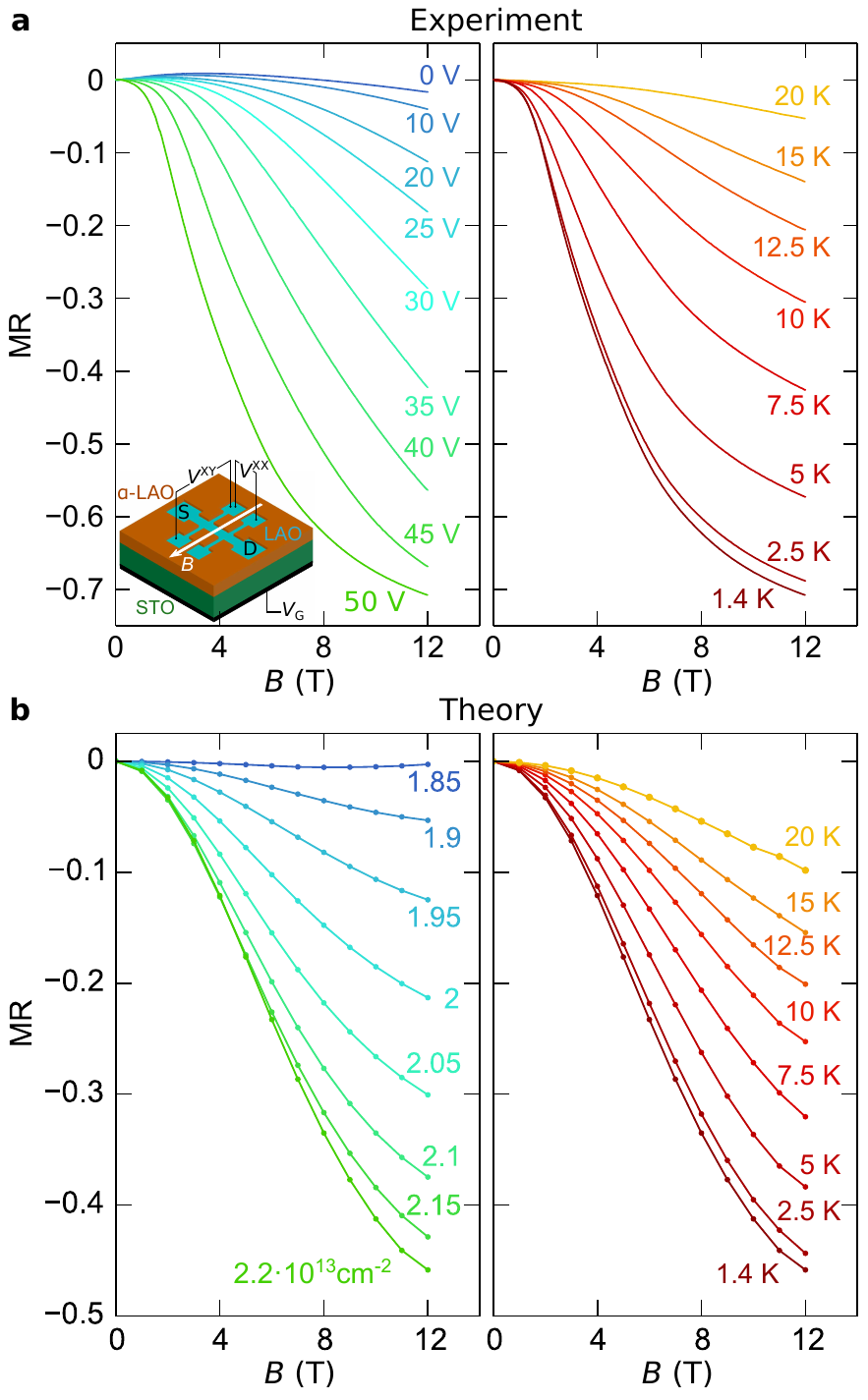}
\caption{(a) Measured magnetoresistance at
$T=1.4\,\si{\kelvin}$ for different gate voltages (left panel) and at
$V_{\mathrm{G}}=50\,\si{\volt}$ for various temperatures (right panel).
Inset: Schematic drawing of the device in a Hall bar geometry (in-plane field perpendicular to current direction),
showing the source S, drain D,
longitudinal voltage $V^\mathrm{{xx}}$, transverse voltage $V^\mathrm{{xy}}$
and gate voltage $V_\mathrm{{G}}$.\\
(b) Magnetoresistance calculated from the Boltzmann equation,
at fixed $T=1.4\,\si{\kelvin}$  (left panel) and at fixed
$n=2.2\cdot 10^{13}\,{\rm cm}^{-2}$ (right panel).}
\label{fig:1}
\end{figure}

We first present the experimental data and then turn to the theoretical description.
Devices were fabricated by using amorphous \ce{LAO} ($\alpha$-\ce{LAO}) as a
hard mask and epitaxially depositing a thin (12\,u.\,c.) film of \ce{LAO}
on top of a \ce{TiO2}-terminated \hkl(001)\ce{STO} single crystal substrate.
The film was grown by pulsed laser deposition at $770\,\si{\celsius}$ in
\ce{O2} at a pressure of $6\cdot 10^{-5}\,\si{\milli\bar}$. The laser fluence
was $1\,\si{\joule\per\centi\meter\squared}$ and
the repetition rate was $1\,\si{\hertz}$. The growth of the film was
monitored \textit{in-situ} using reflection high energy electron
diffraction ({\sc rheed}), and layer-by-layer growth was confirmed. After
deposition, the sample was annealed for one hour at $600\,\si{\celsius}$ in
$300\,\si{\milli\bar}$ of \ce{O2}. Finally, the sample was cooled down to
room temperature in the same atmosphere. Magnetotransport measurements
were performed in a four-probe Hall bar geometry and in a field-effect configuration
(Fig.\,\ref{fig:1}a, inset) established with a homogeneous metallic back gate.
The magnetic field $B$ is applied in-plane and perpendicular to the current. The longitudinal sheet resistance $\rho_{xx}(B)$ determines the dimensionless magnetoresistance
\begin{equation}
\mathrm{MR}(B) = \rho_{{xx}}(B)/\rho_{{xx}}(0)-1. \label{MRdef}
\end{equation}

The left panel of Fig.\,\ref{fig:1}a shows the measured magnetoresistance as a function
of magnetic field, recorded at $1.4\,\si{\kelvin}$, for gate voltages $V_{\rm G}$ ranging from
$0\,\si{\volt}$ to $50\,\si{\volt}$. In general, we observe the magnetoresistance to remain mainly
flat up to some characteristic value of the magnetic field. For larger values,
the magnetoresistance drops sharply. At even higher magnetic fields, the magnetoresistance begins to saturate,
producing an overall bell-like curve.
At the highest voltage $V_{\mathrm{G}}=50\,\si{\volt}$, a very large negative magnetoresistance
is observed (a drop of $70\%$) over a magnetic field range from $0\,\si{\tesla}$ to
$12\,\si{\tesla}$. As $V_{\mathrm{G}}$ is decreased, the overall magnitude of the
magnetoresistance drop is suppressed, as the curves flatten out and the characteristic field
progressively moves to higher $B$. At $V_{\mathrm{G}}=10\,\si{\volt}$, the maximum
magnetoresistance variation is less than 5\%.

The right panel of Fig.\,\ref{fig:1}a shows the measured magnetoresistance at a fixed gate voltage of $\mathrm{V_G}=50\,\si{\volt}$, for
different temperatures ranging from $1.4\,\si{\kelvin}$ to $20\,\si{\kelvin}$.
The correspondence between the bell-shaped magnetoresistance profiles as a function of temperature and
gate voltage is striking. As $T$ increases or $V_{\mathrm{G}}$ decreases,
both the magnitude of the magnetoresistance and steepness of $\partial$MR/$\partial$B decrease.
Although the negative magnetoresistance is progressively suppressed as the temperature is
raised, it is still clearly visible at $20\,\si{\kelvin}$, in agreement with previous experiments \cite{benshalom2009, Wan11}.
Notice that the characteristic field scale of the resistance drop increases with temperature.

For the theoretical description we use a three-band model of the $t_{\rm 2g}$ conduction
electrons at the LAO/STO interface \cite{joshua2012}, with Hamiltonian
\begin{align}
H = \sum_{{\bm k},l, l', \sigma, \sigma'}c_{{\bm k},l,\sigma}^\dagger \left(H_{\rm L}+
H_{\rm SO}+H_{\rm Z}+H_{\rm B}\right)c_{{\bm k},l',\sigma'}.
\label{eq:Hc}
\end{align}
The operators $c_{{\bm k},l,\sigma}^\dagger$ create electrons of spin \(\sigma\)
and momentum \({\bm k}\) (measured in units of the lattice constant $a=0.4\,{\rm nm}$),
in orbitals $l=d_{xy}, d_{xz}, d_{yz}$ of the \ce{Ti} atoms close to
the interface. We describe the various terms in this three-band Hamiltonian, with parameter values from the literature \cite{joshua2012,mattheis1972,santander2011,zhong2013,vanherringen2013,cancellieri2014,satpathy2014,plumb2014,fete2014,salluzzo2009,sm} that we will use in our calculations. (Further details are given in the supplemental material.)

The lobes of the $d_{xy}$
orbital are in-plane, producing two equivalent hopping integrals
$t_l=340\,{\rm meV}$. For the two other orbitals, the $x$-lobe or $y$-lobe is in-plane
and the $z$-lobe is out-of-plane, giving rise to one large and one small
hopping element $t_l$ and $t_h=12.5\,{\rm meV}$, respectively. The $d_{xz}$
and $d_{yz}$ orbitals are hybridized by a diagonal hopping $t_d=t_h$.
Confinement lowers the $d_{xy}$ orbital in energy by
$\Delta_E = 60\,{\rm meV}$. All this information
is encoded in
\begin{align}
&  H_{\rm L} =
  \begin{pmatrix}
	  \epsilon_{xy}(k)- \Delta_{\rm E} & 0 & 0 \\
	 0 & \epsilon_{xz}(k)& \delta(k) \\
	 0 & \delta(k) & \epsilon_{yz}(k)
  \end{pmatrix}\otimes \hat{\sigma}_0,
  \label{eq:HL}\\
\epsilon_{xy}(k)&2t_l(2 -\cos k_x - \cos k_y),\nonumber\\
\epsilon_{xz}(k)&=2t_l(1-\cos k_x) + 2t_h(1-\cos k_y),\\
\epsilon_{yz}(k)&=2t_h(1-\cos k_x) + 2t_l(1-\cos k_y),\nonumber\\
\delta(k)&=2t_d \sin k_x \sin k_y.\nonumber
\end{align}
We use $\hat{\sigma}_{x,y,z}$ and $\hat{\sigma}_0$ to denote the Pauli-matrices and the identity acting on the electron spin.

The intrinsic electric field at the interface breaks inversion
symmetry and produces the term
\begin{align}
  H_{\rm Z} = \Delta_{\rm Z}
  \begin{pmatrix}
	 0 & i\sin k_y & i\sin k_x \\
	 -i\sin k_y & 0 & 0 \\
	 -i\sin k_x & 0 & 0 \\
  \end{pmatrix} \otimes \hat{\sigma}_0,
  \label{eq:HZ}
\end{align}
with $\Delta_Z=15\,{\rm meV}$.
Atomic spin-orbit coupling gives
\begin{align}
H_{\rm SO} = \frac{\Delta_{\rm SO}}{2}
\begin{pmatrix}
0 & i\hat{\sigma}_x & -i\hat{\sigma}_y \\
-i\hat{\sigma}_x & 0 & i\hat{\sigma}_z \\
i\hat{\sigma}_y & -i\hat{\sigma}_z & 0
\end{pmatrix},
\label{eq:HSO}
\end{align}
with an amplitude $\Delta_{\rm SO}=5\,{\rm meV}$.
Together, $H_{\rm Z}$ and $H_{\rm SO}$ cause a Rashba-type
splitting of the bands, coupling the $d_{xy}$ orbital with the
$d_{xz/yz}$ orbitals above the Lifshitz point at the bottom of the $d_{xz/yz}$ bands.

The term \(H_{\rm B}=\mu_{\rm B}(\bm L + g\bm S)\cdot \bm B/\hbar\), with $g=5$ \cite{fete2014},
describes the coupling of the applied magnetic field to the spin and
orbital angular momentum of the electrons, where $\bm S = \hbar\hat{\bm \sigma}/2$ and
\begin{align}
 L_x = \hbar
 \begin{psmallmatrix}
	0 & i & 0 \\
	-i & 0 & 0 \\
	0 & 0 & 0 \\
 \end{psmallmatrix},\,
 L_y = \hbar
 \begin{psmallmatrix}
	0 & 0 & -i \\
	0 & 0 & 0 \\
	i & 0 & 0 \\
 \end{psmallmatrix},\,
 L_z = \hbar
 \begin{psmallmatrix}
	0 & 0 & 0 \\
	0 & 0 & i \\
	0 & -i & 0 \\
 \end{psmallmatrix}.
  \label{eq:L}
\end{align}

\begin{figure}
\includegraphics[width=\linewidth]{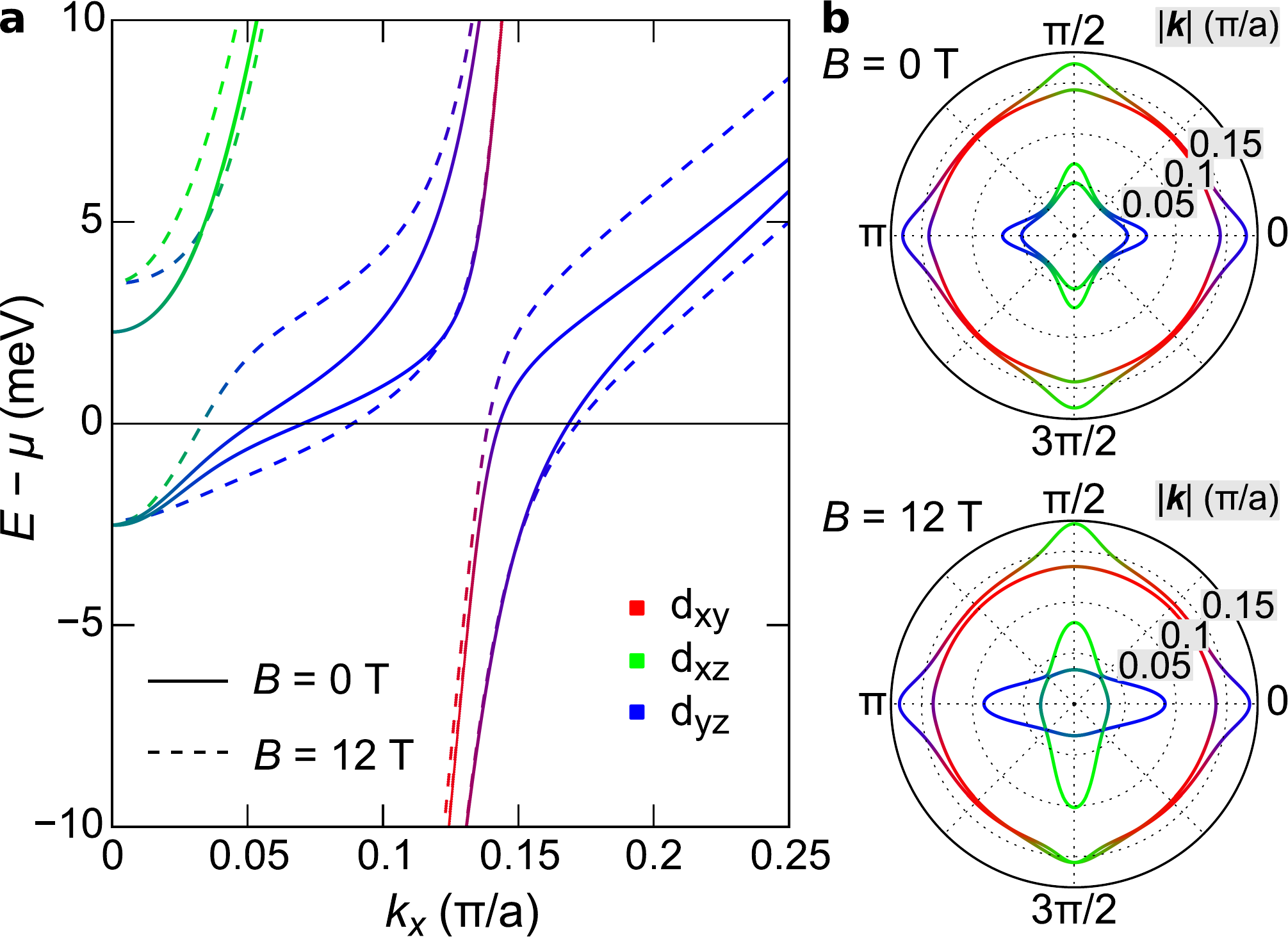}\\
\caption{(a) Dispersion relation for the mobile electrons at the
\ce{LaAlO3}/\ce{SrTiO3} interface, calculated from the model Hamiltonian \eqref{eq:Hc} for $n=2.2\cdot 10^{13}\,{\rm cm}^{-2}$
at $B=0\,\si{\tesla}$ (solid line) and $B=12\,\si{\tesla}$ (dashed line).
Colors indicate the orbital character of the bands.
(b) Corresponding Fermi surfaces when the chemical potential is located at the ``sweet spot'' above the Lifshitz point where the system becomes very sensitve to changes in carrier density and magnetic field.}
\label{fig:2}
\end{figure}

The resulting highly anisotropic band structure is shown in Fig.\ \ref{fig:2}.
Notice the unusually close relevant energy scales:
When measured from the bottom of the upper, anisotropic
bands, the Fermi energy, spin-orbit coupling induced spin-splitting, Zeeman energy ($10\,\si{\tesla}$)
and temperature ($10\,\si{\kelvin}$) all are on the order of $1\,\rm meV$.

We calculate the magnetoresistance from the model Hamiltonian \eqref{eq:Hc} using the semiclassical Boltzmann transport equation for the momentum $\bm{k}$ and band index $\nu$-dependent distribution function $f_{\k,\nu}=f_0(\epsilon_{\k,\nu})+g_{\k,\nu}$. We linearize around the equilibrium Fermi-Dirac distribution $f_0$, at temperature $T$ and chemical potential $\mu$ (determined self-consistently to obtain a prescribed carrier density $n$). In this way we find the conductivity tensor
\begin{equation}
\sigma_{ij}=e\sum_{{\bm k},\nu}(\bm{v}_{{\bm k},\nu})_i\partial g_{{\bm k},\nu}/\partial E_j
\label{conductivity}
\end{equation}
in linear response to the electric field $\bm E$. The longitudinal resistivity $\rho_{xx}$ then follows upon inversion of the $\bm\sigma$-tensor. The band structure determines the velocity ${\bm v}_{\k,\nu}=\hbar^{-1}\nabla_\k\epsilon_{\k,\nu}$, which is not parallel to the momentum $\hbar\k$ because of the anisotropic Fermi surface.

Calculations of this type are routinely simplified using Ziman's relaxation-time approximation \cite{ziman1961,Zim72}, but the combination of finite-range scattering and anisotropic band structure renders this approximation unreliable \cite{Pik11}. We have therefore resorted to a numerical solution of the full partial differential equation:
\begin{align}
&-e({\bm v}_{\k,\nu}\cdot {\bm E})\partial f_0/\partial \epsilon_{\k,\nu}=(e/\hbar)({\bm v}_{\k,\nu}\times {\bm B})\cdot\nabla_\k g_{\k,\nu}\nonumber\\
&\qquad+
\sum_{\k',\nu'}(g_{\k,\nu}-g_{\k',\nu'})q_{\k\nu,\k'\nu'}\delta(\epsilon_{\k,\nu}-\epsilon_{\k',\nu'}).\label{boltzmann}
\end{align}

Elastic impurity scattering enters with a rate
\begin{equation}
  q_{\k\nu,\k'\nu'}=\tfrac{2}{3}\pi^3\hbar^{-1} \delta^2\xi^4 n_{\rm imp}\,e^{-\xi^2|\k-\k'|^2/2} |\langle u_{{\bm k}\nu}|u_{{ \bm k'}\nu'}\rangle|^2.
  \label{eq:qkkp}
\end{equation}
The impurity density \(n_{\rm imp}\) and scattering amplitude \(\delta\) drop out of the magnetoresistance \eqref{MRdef}, so they need not be specified. The scattering potential has correlation length \(\xi\), for which we take $2\,{\rm nm}\simeq 5$~lattice constants, consistent with  experiments on scattering by dislocations \cite{thiel2009}. (We will discuss the role of this finite correlation length later on.) Both intraband and interband scattering are included via the structure factor
$|\langle u_{{\bm k}\nu}|u_{{ \bm k'}\nu'}\rangle|^2$, which takes into account the
finite overlap $\langle \psi_\nu(\bm k)|V(\bm r)|\psi_{\nu'}(\bm k')\rangle$
of the Bloch states $\psi_\nu(\bm k) = u_{{\bm k}\nu}(\bm r) e^{i\bm k\cdot\bm r}$ and
$\psi_{\nu'}(\bm k') = u_{{\bm k'}\nu'}(\bm r) e^{i\bm k'\cdot\bm r}$.
\footnote{In the presence of strong spin-orbit interactions there can be
additional corrections to Eqs.~\eqref{conductivity} and \eqref{boltzmann} \cite{Sinitsyn2006, Nagaosa2010}. We do not consider these here since we have found that they vanish for in-plane fields \cite{sm}.}

\begin{figure}
\includegraphics[width=\linewidth]{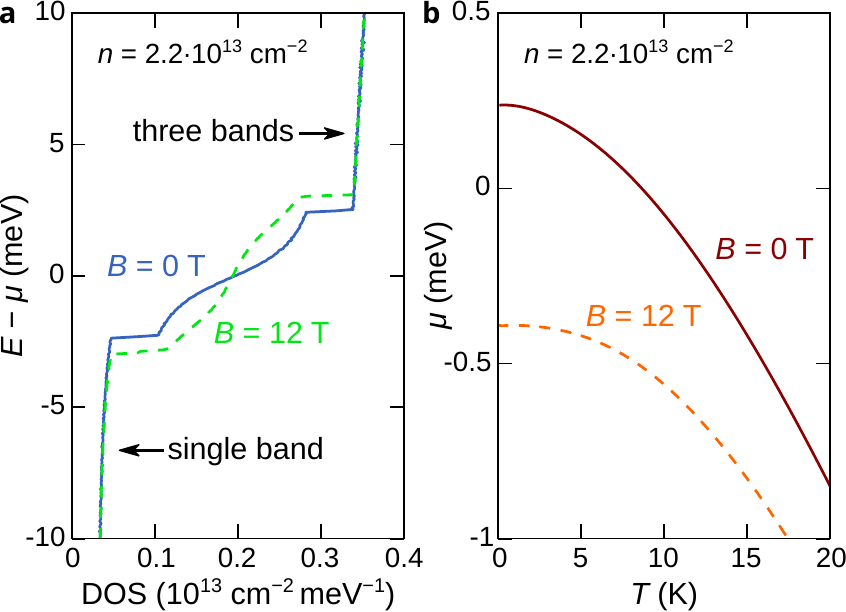}\\
\caption{Energy-dependent density of states (a) and temperature-dependent chemical potential (b),
calculated from the three band Hamiltonian \eqref{eq:Hc}. Both quantities are shown for the ``sweet-spot'' carrier density $n=2.2\cdot10^{13}\,\si{\cm}^{-2}$, at $B=0\,\mathrm{T}$
(solid line) and $B=12\,\mathrm{T}$ (dashed line).}
\label{fig:3}
\end{figure}

The in-plane magnetoresistance resulting from the Boltzmann equation is shown in Fig.\,\ref{fig:1}b. The similarity in the bell-shaped magnetoresistance curves, with a corresponding dependence on carrier density and temperature,
is clear and remarkable in view of the simplicity of the theoretical model. We conclude that a semiclassical single-particle description can produce a ``giant'' magnetoresistance, up to 50\% for a quite conservative choice of parameter values.

Two main ingredients explain how such a large negative magnetoresistance could follow from a model without electron-electron interactions.
The \textit{first ingredient} is the orbital-mixing character of the atomic and inversion-symmetry-breaking spin-orbit coupling terms
$H_{\rm SO}$ and $H_{\rm Z}$.
As a result, the spin-orbit splitting is very non-linear and produces a
``sweet spot'', that is, a narrow range of Fermi energies (carrier densities $n^\ast\simeq 2.2\cdot 10^{13}/{\rm cm}^2$)
in which the system becomes sensitive to small changes in the density.
If the density (or the corresponding gate voltage) is near the sweet spot,
the spin-orbit induced band mixing gives rise to a substantial contribution to the (zero-field) resistance stemming from inter-band scattering.
The Zeeman energy in turn favors an alignment of the spin with the magnetic field and drives a highly anisotropic deformation of the Fermi surface into spin-polarized bands (see Fig.\ \ref{fig:2}).
Inter-band scattering is suppressed which explains the decrease in sheet-resistance.
At densities $n<n^\ast$ only a single band is occupied and spin-orbit coupling is well described by a conventional Rashba term
$\alpha_{\rm SO}(\hat{\bm \sigma}\times\bm p)$\,\cite{caviglia2010, zhong2013, fischer2013} and our calculation gives a vanishingly small magnetoresistance.
At densities $n>n^\ast$ the calculated magnetoresistance starts to saturate and eventually becomes small again.

The \textit{second ingredient} is the finite correlation length $\xi$ of the disorder potential.
The resulting anisotropic scattering rate \eqref{eq:qkkp} is largest at small momenta $|\bm k-\bm k'|$.
Moderate values of $\xi$ on the order of a few lattice constants suppress back-scattering processes within the outer Fermi surface with large average momentum $k_{\rm F}$, while still allowing for inter-band scattering.
This is accompanied by a quasi-particle lifetime which can be significantly smaller for the inner band (smaller average $k_{\rm F}$).
The imbalance of band mobilities promotes the importance of inter-band scattering when transport is dominated by quasi-particles in the outer bands which have a larger Fermi velocity and a small intra-band back-scattering rate.
In comparison we have found \cite{sm} that the isotropic scattering by a delta-function impurity potential cannot produce a magnetoresistance exceeding 15\%.

Our theoretical curves show a smooth dependence on temperature, with the negative magnetoresistance persisting beyond 20~K,
and they show a striking correspondence between the temperature dependence of the magnetoresistance for a
fixed density and the density dependence for a fixed temperature. This correspondence,
a hallmark of our experimental data, can be understood as a consequence
of the renormalization of the chemical potential as a function of
temperature, see Fig.~\ref{fig:3}.
The weak temperature dependence of the Hall resistance point towards a constant carrier density in the range 1--20~K \cite{thiel2006}.
As shown in Fig~\ref{fig:3}a the density of states increases steeply with band energy in the vicinity of the sweet spot, much more than in conventional semiconductors.
To keep the total carrier density fixed with increasing temperature,
the chemical potential is lowered by more than $1\,{\rm meV}$ at $20\,\si{\kelvin}$ compared to its low temperature limit.
This is why increasing the temperature is equivalent
to probing the band structure at a lower energy, explaining the similarity in the magnetoresistance curves in the left and right panels of Fig.~\ref{fig:1}.

\begin{figure}[tb]
  \begin{center}
	 \includegraphics[width=\linewidth]{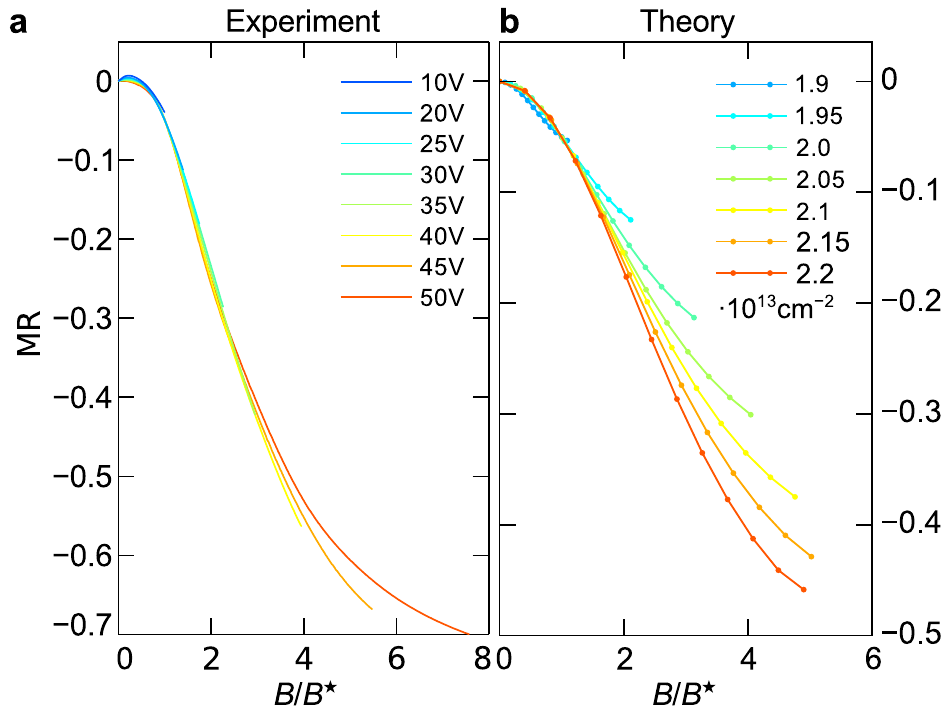}
  \end{center}
  \caption{Measured (a) and calculated (b) magnetoresistance at $1.4\,{\si\kelvin}$ for different densities or gate voltages as a function of the rescaled magnetic field $B/B^\star$. The characteristic field $B^\star$ is chosen such that the rescaled curves all pass through the point with ${\rm MR}=-0.05$.}
  \label{fig:4}
\end{figure}

These are the two key arguments in favor of a single-particle spin-orbit-coupling based mechanism for the giant negative magnetoresistance: Firstly, the persistence of the effect to elevated temperatures, and secondly the corresponding effect of temperature-increase and density-decrease. It seems difficult to incorporate these features of the data in the correlated-electron mechanism \cite{joshua2013,ruhman2014}, based on Kondo-screening of magnetic moments. There is a third noteworthy feature of the data that is not well reproduced by our calculation, and has been interpreted as evidence for a transition into a low-field Kondo phase \cite{joshua2013, ruhman2014}: A rescaling of the magnetic field $B\rightarrow B/B^\star$ by a density-dependent value $B^\star$ collapses the measured magnetoresistance at different densities onto a single curve, see Fig.~\ref{fig:4}a. If we apply this $B/B^\star$ scaling to our numerical results a significant $n$-dependence remains, see Fig.~\ref{fig:4}b. The experimental scaling law points to some relevant physics that is not yet included in our minimal model.

In conclusion, we have presented experimental data and theoretical calculations that support a semiclassical single-particle mechanism for the giant magnetoresistance of the LAO/STO interface. The Boltzman transport equation with spin-orbit coupling, in combination with anisotropy of Fermi surface and scattering rates, suffices to produce a large resistance drop upon application of a magnetic field.  The characteristic temperature and carrier-density dependence agree quite well with what is observed experimentally, but the $B/B^\star$ scaling will likely require an extension of the simplest three-band model.

Our explanation of the sudden onset of the magnetoresistance when the carrier density approaches a ``sweet spot'' of amplified spin-orbit coupling has addressed the normal-state transport above the superconducting transition temperature. Since superconductivity happens in the vicinity of the same ``sweet spot'', it would be interesting to investigate whether spin-orbit coupling plays a dominant role in that transition as well.

\begin{acknowledgements}
We thank A.\,R.~Akhmerov, C. Morais Smith, D.\,I.~Pikulin, J. Ruhmann, and M.~Wimmer for fruitful discussions, Srijit Goswami for support in sample fabrication, and R.~Hoogerheide and M.~van Oossanen for technical support. This work is part of the DITP and Nanofront consortia, funded by the Netherlands Science Foundation NWO/OCW. We acknowledge support by the Academy of Finland Center of Excellence program, the European Research Council (Grant No.\ 240362--Heattronics and a Synergy grant), and the Foundation for Fundamental Research on Matter (FOM).
\end{acknowledgements}

\clearpage
\newpage

\appendix

\renewcommand{\thefigure}{S\arabic{figure}}
\setcounter{figure}{0}

\section*{Supplemental Material}


\subsection{1. Complete set of experimental data}\label{sm:exp_data}

\begin{figure}[tb]
  \begin{center}
	 \includegraphics[width=0.95\linewidth]{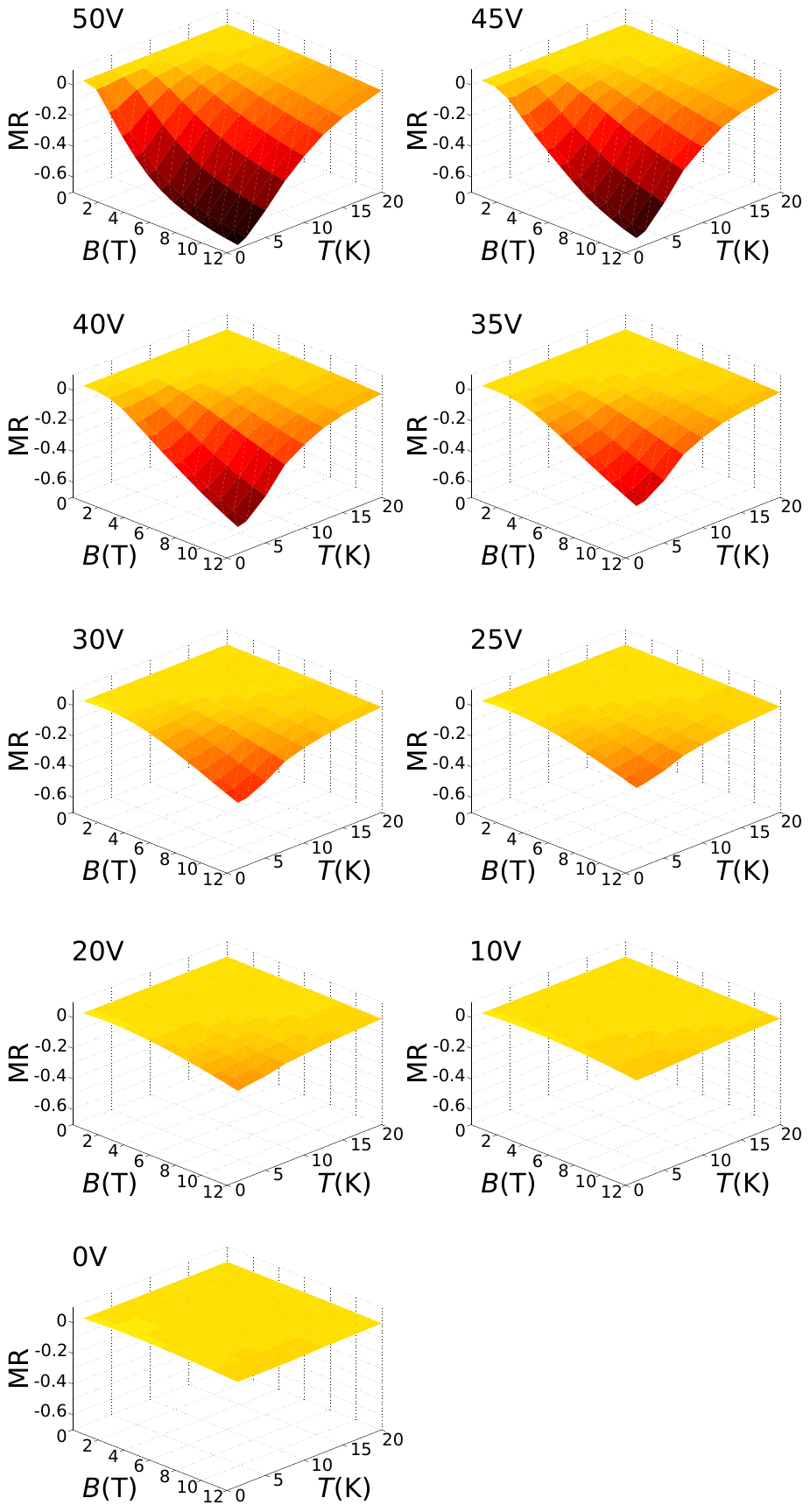}
  \end{center}
  \caption{Measured magnetoresistance in the $B-T$ space of parameters for gate voltages ranging from $50\,\si{\volt}$ to $0\,\si{\volt}$. }
  \label{fig:S2}
\end{figure}

For completeness, Fig.~\ref{fig:S2} shows the systematic study of magnetoresistance in the $B-T$ space of parameters for gate voltages ranging from $50\,\si{\volt}$ to $0\,\si{\volt}$. At high gate voltage and low temperature, a very large negative magnetoresistance is observed (up to 70\% over a magnetic field range of $12\,\si{\tesla}$). As gate voltage (temperature) is decreased (increased), the overall magnitude of the magnetoresistance drop is suppressed, as the curves flatten out.

\subsection{2. Details on the choice of the model parameters}\label{sm:parameters}
Values for the three-band model parameters found in the theoretical literature vary over a wide range, see for example Refs.\ \onlinecite{joshua2012,zhong2013,vanherringen2013}.
{\sc arpes} measurements on the surface of STO \cite{santander2011,plumb2014} and LAO/STO \cite{cancellieri2014} have extracted values for the light and heavy effective masses as well as the confinement splitting $\Delta_{\rm E}$.
The values are similar in all of the experiments.  We take $t_l$, $t_h$ according to the effective masses for the $d_{xy}$ and $d_{xz}/d_{yz}$ band in Ref.~\onlinecite{plumb2014} and $\Delta_E$ according to the value found in Ref.~\onlinecite{santander2011}. An exact determination of the spin-orbit energies $\Delta_{\rm SO}$ and $\Delta_{\rm Z}$ is not yet available experimentally.
There are, however, clear indications that the spin-orbit energy scale may be above $10\,{\rm meV}$ \cite{shalom2010,caviglia2010}.
We take moderate values consistent with the theoretical literature \cite{mattheis1972,zhong2013}.
We note that our simulations suggest that experiments are in the regime $\Delta_{\rm Z}>\Delta_{\rm SO}$. The calculated magnetoresistance is negative in this regime, while we have found both negative and positive magnetoresistance, depending on the density, for $\Delta_{\rm SO}>\Delta_{\rm Z}$.

\subsection{3. Estimate of the ``sweet-spot'' carrier density and the magnetic field sensitive density window.}

\begin{figure}[tb]
  \begin{center}
	 \includegraphics[width=0.75\linewidth]{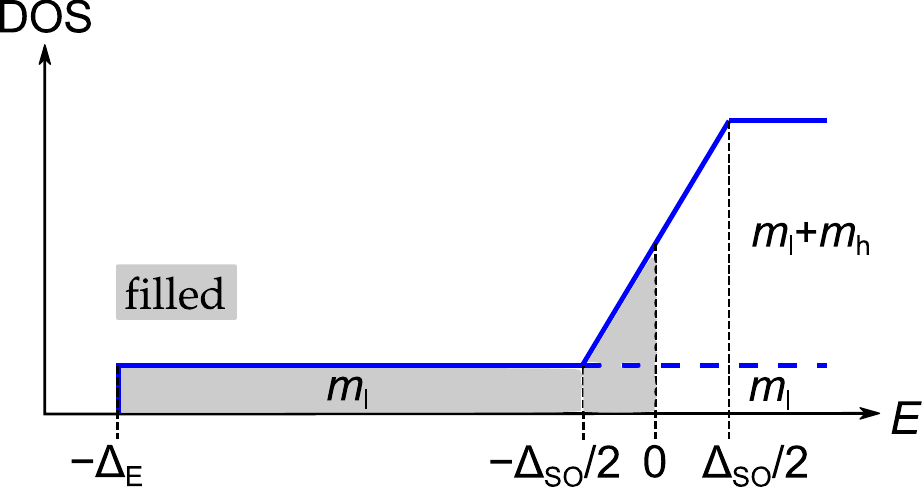}
  \end{center}
  \caption{Approximate density of states and filling for the ``sweet-spot'' carrier density at $T=0$. $m_l=\hbar^2/(2t_la^2)$ and $m_h=\hbar^2/(2t_ha^2)$ refer to the effective light and heavy electron masses corresponding to a 2d density of states $m_i/\pi\hbar^2= 1/2\pi t_ia^2$ for $i=l,h$.}
  \label{fig:S3}
\end{figure}

A central quantity of our proposed model is the ``sweet-spot'' carrier density $n^*$.
This density corresponds to a position of the Fermi level $E_{\rm F}=\mu(T=0)\approx 0$ where band structure is most sensitive to the competition between spin-orbit coupling and magnetic field.
Even for the mimimal three-band model of Hamiltonian~\eqref{eq:Hc} the exact value of $n^*$ has a complicated dependence on the model paramters $t_l$, $t_h$, $t_d$, $\Delta_{\rm E}$, $\Delta_{\rm SO}$ and $\Delta_{\rm Z}$ that can be obtained from integrating the density of states shown in Fig.~\ref{fig:3}.
In order to give simple estimate we can approximate the density of states as shown Fig.~\ref{fig:S3}.
The corresponding estimate for the sweet-spot density is given by
\begin{align}
  n^*a^2 \approx \frac{\Delta_{\rm E}}{2\pi t_l}+\frac{\Delta_{\rm SO}}{16\pi}\left(\frac{1}{t_l}+\frac{1}{t_h}\right).
  \label{eq:nstarestimate}
\end{align}
For our choice of parameters $t_l=340\,{\rm meV}$, $t_h=12.5\,{\rm meV}$, $\Delta_{\rm E}=60\,{\rm meV}$ and $\Delta_{\rm SO}=5\,{\rm meV}$ we obtain $n^*a^2\approx 0.036$.
Numerical integeration of the density of states yields $n^*a^2=0.035$.

Due to the fixed in-plane configuration of the applied magnetic field, the carrier densities corresponding to each measured gate voltage in Fig.~\ref{fig:2}a) could not be determined for this sample.
(Switching from an in-plane to an out-of-plane Hall configuration required a thermal cycling of the device rendering to obtained Hall densities unreliable.)
However, in previous samples with similar geometry and grown under the same conditions, the carrier density modulation resulting from field effect between $0\,\si V$ and $50\,\si V$ is about $0.5\cdot10^{13}\,{\rm cm}^{-2}$ \cite{caviglia2008}, in good agreement with the carrier density values in our model calculation. The density window for which our minimal model shows a large magnetoresistance is essentially limited to a chemical potential window $\Delta\mu\sim\Delta_{\rm SO}$ around the ``sweet-spot'' density.
Both inelastic scattering processes and the presence of additional subbands may extend this energy window, if we go beyond our minimal model.

\subsection{4. Theoretical magnetoresistance for point-like and non-Gaussian scatterers}\label{sm:nonGaussian}
\begin{figure}[tb]
  \begin{center}
	 \includegraphics[width=\linewidth]{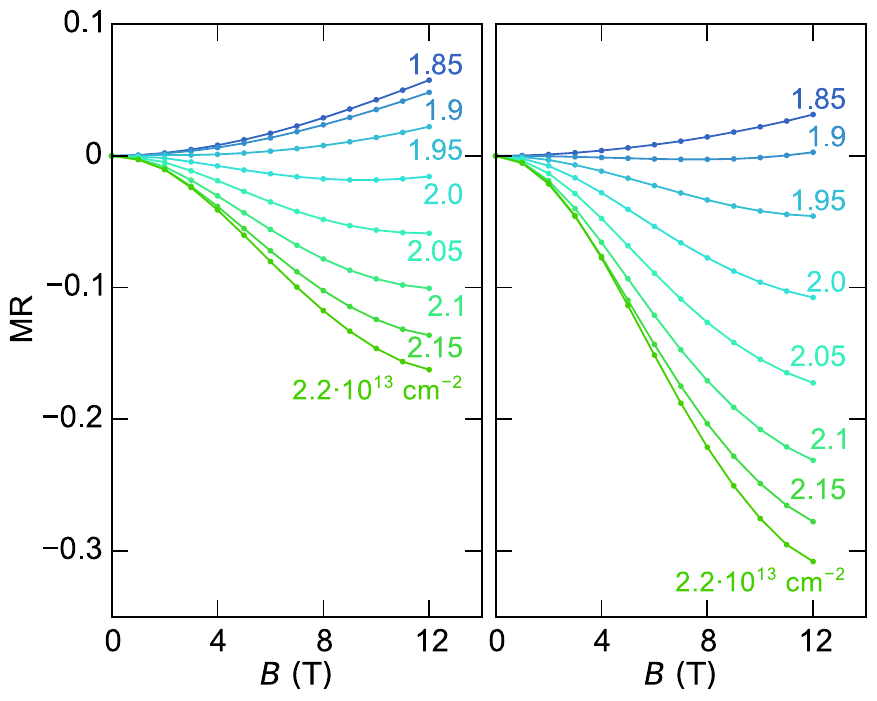}
  \end{center}
  \caption{Theoretical magnetoresistance for the same model parameters as in Fig.~\ref{fig:1}b, but using point-like uncorrelated disorder $q_{{\bm k}\nu,{\bm k'}\nu'}={\rm const.}|\langle u_{{\bm k}\nu}|u_{{\bm k'}\nu'}\rangle|^2$ (or $\xi=0$) (left panel) and same parameters as in Fig.~\ref{fig:1}b but using $q_{{\bm k}\nu,{\bm k'}\nu'}\propto |\langle u_{{\bm k}\nu}|u_{{\bm k'}\nu'}\rangle|^2(|\bm k-\bm k'|^2 + 1/l^2)^{-1}$ for $l=5$ lattice constants (right panel).}
  \label{fig:S1}
\end{figure}
In the main text we discussed how the amplitude of the calculated magnetoresistance drop is larger for disorder with a finite correlation length $\xi>0$. For comparison, we show in the left panel of Fig.~\ref{fig:S1} the magnetoresistance for the same parameters as in the main text, but point-like, uncorrelated scatterers. Notice that the maximum drop in this case is only about $15\%$, more than a factor of 3 smaller. Moreover the magnetoresistance is actually positive for a range of densities above the Lifshitz-point ($n_{\rm L}=1.83\cdot10^{13}/{\rm cm}^2$), but below  the sweet-spot density $n^\ast=2.2\cdot10^{13}/{\rm cm}^2$.

While it is important that the scattering amplitude has a finite correlation length, it need not necessarily be a Gaussian correlation. For comparison in the right panel of Fig.~\ref{fig:S1} we show results for a scattering amplitude proportional to $(|\bm k-\bm k'|^2 + 1/l^2)^{-1}$, like it might be produced by screened Coulomb potentials of charged impurities close to the interface. Contrary to the Gaussian case there is now a significant amount of scattering at large momenta including backscattering. Still we find that for a screening length $l$ of 5 lattice constants the magnetoresistance is already enhanced by a factor of 2 compared to point-like scatterers and the positive magnetoresistance at lower densities is suppressed.
\subsection{5. Theoretical magnetoresistance as a function of the alignment between the magnetic field and the plane of the 2DES}
\begin{figure}[tb]
  \begin{center}
	 \includegraphics[width=\linewidth]{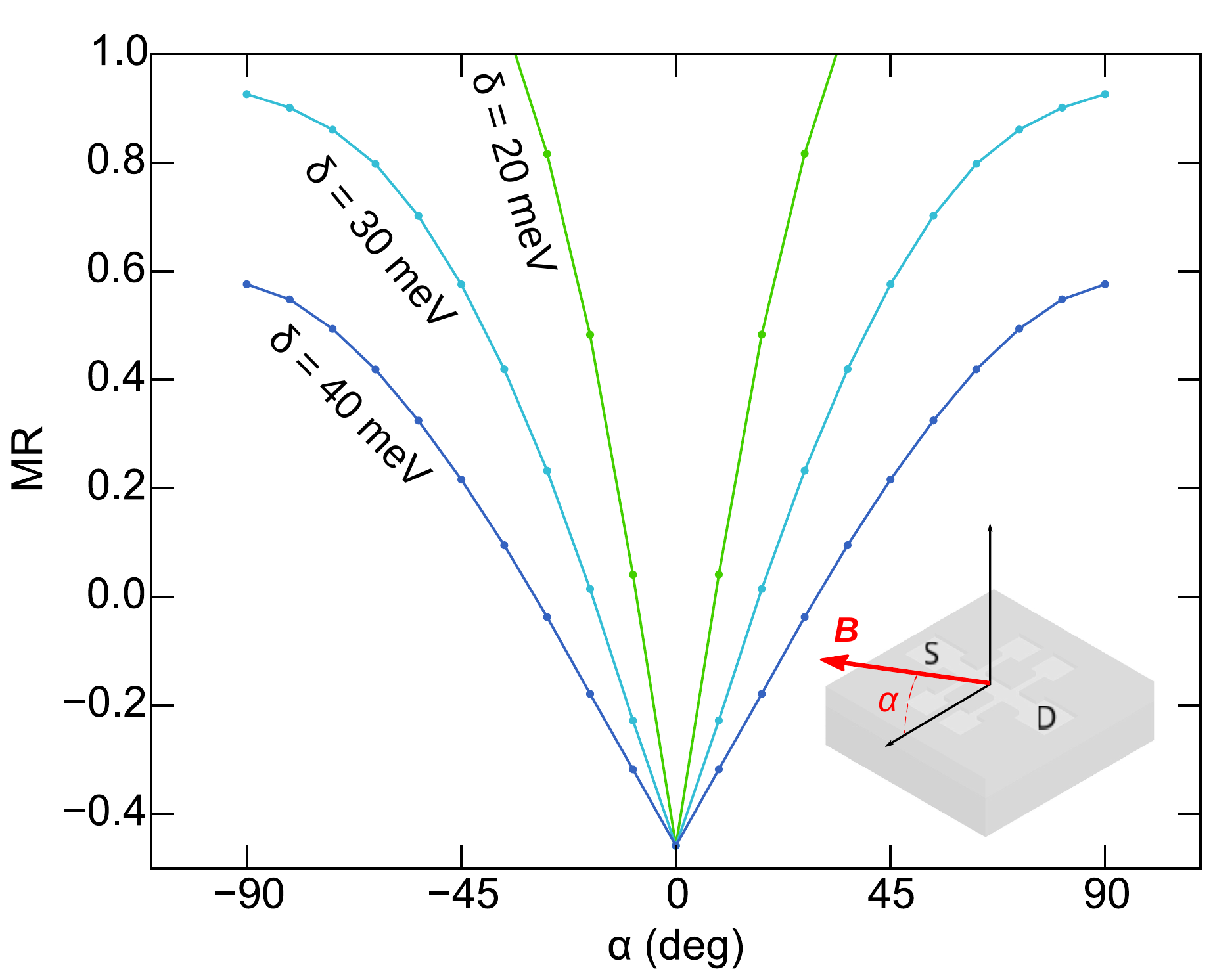}
  \end{center}
  \caption{Calculated magnetoresistance as a function of the angle $\alpha$ (see inset) between the magnetic field $\bm B$ and the plane of the 2DES. Shown is the dependence for a fixed amplitude $|\bm B|=12\,\si{T}$ at $1.4\,{\si K}$ for the ``sweet-spot'' carrier density $n=2.2\cdot 10^{13}\,{\rm cm}^{-2}$ using the same band parameters as in Fig.~\ref{fig:2}b. In the case of an out-of-plane field the magnetoresistance explicitly depends on the overall scattering amplitude $\delta^2n_{\rm imp}$. Here we show three examples for the same impurity density $n_{\rm imp}=1/(5\xi)=1/(25a^2)$. The disorder amplitude $\delta$ is chosen such that the calculated zero-field resistance is comparable with our experimental values at high gate voltages, $\rho_0=70\,{\si\ohm}$, $160\,{\si\ohm}$, $275\,{\si\ohm}$ for $\delta=20\,{\rm meV}$, $30\,{\rm meV}$, $40\,{\rm meV}$, respectively.}
  \label{fig:S4}
\end{figure}
Previous experiments (see for example Ref.\ \onlinecite{benshalom2009}) have shown that the giant magnetoresistance has a strong out-of-plane anisotropy.
Upon application of an out-of-plane component to the magnetic-field the negative magnetoresistance quickly decays and turns positive.
For a comparison of our model with this characteristic experimental feature we show the calculated magnetoresistance as a function of the angle $\alpha$ between the magnetic field $|\bm B|$ and the plane of the 2DES for the ``sweet-spot'' carrier density $n^*$ and the same parameters as Fig.~\ref{fig:2}b), left panel, see Fig.~\ref{fig:S4}.
Our minimal reproduces the general shape of the observed out-of-plane anisotropy.
There is a sharp dip for perfect in-plane alignment.
Upon application of an out-of-plane component the negative magnetoresistance signal becomes positive.
This anisotropy is a consequence of the planar anisotropy of the spin-orbit coupling in the 3-band Hamiltonian, as well as the absence of orbital effects $(\bm B \times \nabla_{\bm k})$ for the in-plane fields.
The resulting magnetoresistance dip explicitly depends on the overall scattering amplitude $\delta^2n_{\rm imp}$.
Fig.~\ref{fig:S4} shows three examples.
The dip is sharpest for small values of the disorder amplitude $\delta$.
We note that the longitudinal resistance obtained from the Boltzmann equation at a $12\,{\si\tesla}$ perpendicular field is likely to be an overestimate, because at large perpendicular fields additional ``skipping''-orbit channels appear from that are missing in the Boltzmann approach.
These would cause an increase of the Hall conductivity $\sigma_{xy}$ and a decrease in the longitudinal resistance.

\subsection{6. Spin-orbit corrections to the Boltzmann transport}
When the wave functions of the conducting electrons have a non-trivial orbital and spin character, like in
multiband spin-orbit coupled materials, new intrinsic and extrinsic mechanisms that are absent in simpler systems
show up in determining the transport properties of complex materials.
Three different mechanisms that are well known studied in the framework of the anomalous Hall effect can be discussed and systematically included in the Boltzmann transport description\cite{Nagaosa2010}.

The first correction to Eqs.~\eqref{conductivity} and \eqref{boltzmann} is not dependent on the scattering process, because it
follows from an intrinsic property of the band structure.
The non-trivial Berry curvature $F_{\k,\nu}$ of the bands in presence of spin-orbit coupling
acts as a magnetic field in the momentum space and couples
to the electric field to give an additional velocity $-{\bm F}_{\k,\nu} \times e{\mathbf{E}}$ to the quasi-particle in the state
$(\k, \nu)$.
As said before the topological correction does not depend on the details of the scattering. So it can be dominant
or subdominant (with respect to the mechanisms we discuss below) depending on the strength of the disorder and the density of impurities.
However, its contribution to the (transversal) conductivity always remains of the order of the conductance quantum.
We find that both measured and calculated longitudinal conductivities are much larger than $e^2/h$.
In comparison with the longitudinal magnetoresistance which is of the order of the total resistance it can thus only give rise to small corrections.
We do not consider these here.

The two mechanisms that we discuss below instead explicitly depend on the scattering processes.
The first correction originates from asymmetric (\emph{skew}) scattering of polarized electrons accelerated by an electric field.
As we do in the main text impurity scattering is commonly treated in the lowest Born approximation, where the transition rate
is given by the Fermi golden rule $q_{\k'\nu', \k \nu} = 2\pi {\vert \langle \k' \nu' | V | \k \nu \rangle \vert}^2/ \hbar$.
One of the limitations of this approximation is that it does not take into account the skew scattering, because ${\vert V_{\k'\nu', \k \nu}\vert}^2$
is clearly symmetric upon exchange of the initial and final states.
In order to include antisymmetric corrections, the rate
must be computed including higher orders in the perturbative expansion of the full scattering $T$-matrix.
The first skew term is proportional to $V^3$.
At this order,
the semiclassical equation is still fully consistent compared to a rigorous quantum mechanical calculation.
In the weak disorder limit (to linear order in the impurity density $n_{\rm imp}$) the antisymmetric component of the transition probability is
\begin{eqnarray}
 q^{sk}_{\k' \nu', \k \nu}=
 \quad-\frac{(2\pi)^2}{\hbar} \sum_{\q, \nu''} &\Bigg[\Im \Big( \langle
 V_{\k'\nu', \k \nu} V_{\k \nu, \q \nu''} V_{\q \nu'', \k' \nu'} \rangle_{dis} \Big ) \nonumber\\
 &\cdot\delta (\epsilon_{\k,\nu} - \epsilon_{\q,\nu''})\Bigg],
\end{eqnarray}
where
\begin{eqnarray}
 &V_{\k'\nu', \k \nu} V_{\k \nu, \q \nu''} V_{\q \nu'', \k' \nu'} \propto \nonumber\\
 &\quad\langle u_{\k' \nu'} \vert u_{\k \nu}\rangle \langle u_{\k \nu} \vert u_{\q \nu''}\rangle
 \langle u_{\q \nu''} \vert u_{\k' \nu'}\rangle
 \label{skew}
\end{eqnarray}
where the disorder average has been introduced.
Notice that naturally $q^{sk}$ violates the detailed-balance condition, but still an important sum-rule is satisfied.
\begin{eqnarray}
&q^{sk}_{\k'\nu', \k \nu}
= -q^{sk}_{\k \nu, \k'\nu'},\\
&\sum_{\k', \nu'} q^{sk}_{\k' \nu', \k \nu} =
\sum_{\k', \nu} q^{sk}_{\k \nu, \k' \nu'} = 0.
\end{eqnarray}
We do not consider further contributions on the order $n_{\rm imp}^2$, here.

In addition to skew-scattering, the electronic wave packet accelerated by an electric field is subjected to the shift $\delta {\bm r}_{\k',\k}$ (\emph{side-jump})
of its center of mass during a scattering event. The gauge-invariant expression
for the coordinate shift can be expressed in terms of the Pancharatnam-Berry phase $\Phi_{\q \nu'',\k \nu,\k' \nu'}$ \cite{Sinitsyn2006},
\begin{align}
\delta {\bm r}_{\k' \nu',\k \nu} = &- \Big( \frac{\partial}{\partial \k''} \Phi_{\q \nu'',\k \nu,\k' \nu'} \Big)_{\k'' \rightarrow \k} \nonumber\\
 &- \Big( \frac{\partial}{\partial \k''} \Phi_{\q \nu'',\k \nu,\k' \nu'} \Big)_{\k'' \rightarrow \k'},
 \label{side-jump}
\end{align}
\begin{align}
 \Phi_{\q \nu'',\k \nu,\k' \nu'}= &{\rm arg}\left(\langle u_{\q \nu''} \vert u_{\k \nu}\rangle \langle u_{\k \nu} \vert u_{\k' \nu'}\rangle
 \langle u_{\k' \nu'} \vert u_{\q \nu''}\rangle\right) .
\end{align}
The presence of the side-jump has two effects on the transport.
First, the accumulation of coordinate shifts after many scattering events
gives (in the lowest Born approximation) a correction to the velocity
${\bm v}^{sj}_{\k \nu} = \sum_{\k', \nu'} q_{\k' \nu', \k \nu}
\delta {\bm r}_{\k'\nu',\k\nu}$.
Second, a particle scattered by an impurity under side-jump acquires a kinetic energy
$\Delta \epsilon_{\k' \nu', \k \nu} = e{\bm E} \cdot \delta {\bm r}_{\k' \nu',\k \nu}$ in order to compensate the change
in the potential energy induced by the electric field. As a consequence, the equilibrium distribution $f_0$ experiences an additional shift:
\begin{eqnarray}
 f_0(\epsilon_{\k,\nu})-f_0(\epsilon_{\k',\nu'}) =
 -(\partial f_0 / \partial \epsilon_{\k,\nu}) \Delta \epsilon_{k' \nu', \k \nu}\ .
\end{eqnarray}
Including all the terms, the conductivity tensor is given by
\begin{equation}
  \sigma_{ij}=e\sum_{{\bm k},\nu}(\frac{1}{\hbar}\frac{\partial \epsilon_{\k, \nu}}{\partial \k} + v^{sj}_{\k, \nu}
  )_i\partial g_{{\bm k},\nu}/\partial E_j -\varepsilon_{ij} eF^z_{\bm k,\nu}f_0(\epsilon_{\bm k,\nu}),
\end{equation}
where $\varepsilon_{ij}$ is the two-dimensional antisymmetric tensor and $g_{{\bm k},\nu}$ solves the modified Boltzmann equation \cite{Nagaosa2010}
\begin{widetext}
\begin{align}
  -e\Big(\frac{1}{\hbar}\frac{\partial \epsilon_{\k, \nu}}{\partial \k}\cdot {\bm E}\Big)\frac{\partial f_0}{\partial \epsilon_{\k,\nu}}=
\sum_{\k',\nu'} \Big( q_{\k'\nu',\k\nu}+q^{sk}_{\k'\nu',\k\nu} \Big)
\Big (g_{\k,\nu}-g_{\k',\nu'}- \frac{\partial f_0}{\partial \epsilon_{\k,\nu}}
\Delta \epsilon_{\k' \nu', \k \nu} \Big )\delta(\epsilon_{\k,\nu}-\epsilon_{\k',\nu'}) .
&\label{boltzmann_skew}
\end{align}
\end{widetext}

Discarding the intrinsic Berry-curvature correction, we numerically solve equation \eqref{boltzmann_skew} for scattering from correlated impurities where the amplitude of the disorder potential is
uniformly distributed in the asymmetric range $\delta[-1+\Delta,1+\Delta]$ (notice that for a symmetric distribution the Gaussian
correlator $\langle V^3\rangle_{dis}$ appearing in the skew-scattering term automatically vanishes).
However, we find that the product of the three overlaps $\langle u_{\q \nu''} \vert u_{\k \nu}\rangle \langle u_{\k \nu} \vert u_{\k' \nu'}\rangle
 \langle u_{\k' \nu'} \vert u_{\q \nu''}\rangle$ is strictly real for arbitrary momenta and band indices when the magnetic field is applied in the
 plane of the 2DES. Hence both the skew-scattering \eqref{skew} and the side-jump \eqref{side-jump} terms turn out to be zero
 for in-plane field.

Although here we computed the Pancharatnam-Berry phase numerically, it is easy to analytically show the same result but for the simpler case of
the Rashba Hamiltonian.
In momentum space, the Rashba Hamiltonian has a 2x2 matrix structure.
Hence the relevant product of wavefunction overlaps, or more precisely, the argument of this quantity (the Pancharatnam-Berry phase) is equivalent to half the solid angle the Bloch
states $u_{\bm k}$, $u_{\bm k'}$, and $u_\q$ span on the Bloch sphere.
Since for an in-plane Zeeman term the Rashba Hamiltonian may be expanded solely in terms of the
Pauli matrices $\hat{\sigma}_0$, $\hat{\sigma}_x$, and $\hat{\sigma}_y$,
this solid angle vanishes identically. The same phenomenon leads to the vanishing of the side-jump.

If an out-of-plane magnetic field is switched on, all the contributions become finite, but we find them to remain small throughout our simulations.
More explicitly, for the same choice of parameters of the calculations in the main text we observe numerically that both skew-scattering and side-jump contributions yield corrections less than 1\% of the calculated total magnetoresistance resistance for out-of-plane fields up to $12\,{\si\tesla}$ and distributions as asymmetric as $\Delta = 0.5$.
We do not show these results here as they are almost invisible on the scale of Fig.~\ref{fig:S4}.
\end{document}